\documentclass[12pt]{article}
\usepackage{mathrsfs}
\usepackage{amsmath,amssymb}

\usepackage{amsfonts}
\usepackage{latexsym}
\usepackage{amsthm}
\usepackage{pictex}

\usepackage[T2A]{fontenc}     
\usepackage[cp1251]{inputenc} 
\usepackage{graphicx}

\def\neigh{neighborhood{}}

\def\Re{{\rm Re\,}}
\def\Im{{\rm Im\,}}
\def\lb{\label}
\def\dSRN{de Sitter-Reissner-Nordstr{\"o}m }
\def\dSS{de Sitter-Schwarzschild }
\newtheorem{theorem}{Theorem}

\newtheorem{lemma}{Lemma}

\def\a{\alpha}

\def\G{\Gamma} \def\cD{{\cal D}}

  \def\cH{{\cal H}}   \def\mH{{\mathscr H}}
   \def\cI{{\cal I}}

 \def\cL{{\cal L}}   
\def\l{\lambda}   
\def\L{\Lambda}   
       
    \def\cP{{\cal P}}   
      
\def\s{\sigma} \def\cR{{\cal R}}   
\def\S{\Sigma}

\def\o{\omega}

\def\mH{{\mathscr H}}

\def\R{{\Bbb R}}
\def\C{{\Bbb C}}

\def\N{{\Bbb N}}
\def\S{{\Bbb S}}

\def\qq{\quad}
\newcommand{\ma}{\begin{pmatrix}}
\newcommand{\am}{\end{pmatrix}}
\newcommand{\ca}{\begin{cases}}
\newcommand{\ac}{\end{cases}}

\let\geq\geqslant
\let\leq\leqslant
\def\ma{\left(\begin{array}{cc}}
\def\am{\end{array}\right)}

\let\geq\geqslant
\let\leq\leqslant
\def\[{\begin{equation}}
\def\]{\end{equation}}

\def\no{\noindent}

\def\/{\over}

\def\os{\oplus}

\def\Re{\mathop{\rm Re}\nolimits}
\def\Im{\mathop{\rm Im}\nolimits}

\def\BBox{\hspace{1mm}\vrule height6pt width5.5pt depth0pt \hspace{6pt}}

\begin{document}

\date{\today}

\title{Resonance expansions of massless Dirac fields propagating in the exterior  of a  de~Sitter-Reissner-Nordstr{\"o}m black hole.}

\author{
Alexei Iantchenko
\begin{footnote}
{Department of Materials Science and Applied Mathematics, Faculty of Technology and Society, Malm\"{o} University, SE-205 06 Malm\"{o}, Sweden, email: ai@mah.se }
\end{footnote}
}

\maketitle

\begin{abstract}
We give an expansion of the solution of the evolution equation for the massless Dirac fields in the outer region of de Sitter-Reissner-Nordstr\"om black hole in terms of resonances. By means of this method we  describe the decay of local energy for compactly supported data.
The proof uses the cut-off resolvent estimates for the semi-classical Schr\"odinger operators from  \cite {BonyHafner2008}. The method extends to the Dirac operators on spherically symmetric asymptotically hyperbolic manifolds.

\end{abstract}


\section{Introduction and main results.}

Quasi-normal modes (QNM) are well known to play an
important role in black hole physics. They determine the
late-time evolution of fields in the black hole exterior and
 eventually dominate
the black hole response to any kind of perturbation. 

QNMs of a black hole are defined as proper solutions of the perturbation equations belonging to certain complex characteristic frequencies
(resonances) which satisfy the boundary conditions appropriate for purely ingoing waves at the event horizon and purely outgoing waves at infinity (see \cite{ChandrasekharDetweller1975}, \cite{Chandrasekhar1983}).
For the physics review we refer to \cite{KokkotasSchmidt1999} and more recent \cite{Bertietal2009}.

From quasi-normal frequencies one can extract information of the physical parameters of the black hole - mass, electric charge, and angular momentum - from the gravitational wave signal by fitting the observed quasinormal frequencies to those predicted from the mathematical analysis.

The subject has become very popular for the last few decades including the development of   stringent mathematical theory of QNMs (see  \cite{BachelotMotetBachelot1993}, \cite{SaBarretoZworski1997}, \cite{DafermosRodnianski2007}, \cite{Dyatlov2011}, \cite{Dyatlov2012}, \cite{Gannot2014}).

The paper  of S{\'a} Barreto and Zworski \cite{SaBarretoZworski1997} provides with stringent mathematical justification for  localization of QNMs for the wave equation on the
de Sitter-Schwarzschild metric. In Regge-Wheeler coordinates the problem is reduced to the scattering problem for the  Schr{\"o}dinger equation on the line with exponentially decreasing potential.
In the  Schwarzschild  case (zero cosmological constant, which corresponds to asymptotically flat Universe) the Regge-Wheeler potential is only polynomially decreasing and the method does not work due to the possible accumulation of resonances at the origin. A non-zero cosmological constant is needed in order to apply results of \cite{MazzeoMelrose1987} and \cite{Guillarmou2005}, and to define an analytic continuation of the resolvent in a proper space of distributions.

Later, work \cite{SaBarretoZworski1997} was complemented by the paper of Bony and H{\"a}fner \cite{BonyHafner2008}, where the authors considered the local energy decay for the wave equation on the
de Sitter-Schwarzschild metric and proved expansion of the solution  in terms of resonances.

The works  
 \cite{Dyatlov2011} and \cite{Dyatlov2012} of Dyatlov provided with  detailed analysis of QNMs for
more realistic and difficult rotating Kerr-de Sitter black holes. The situation was complicated by the absence of spherical symmetry. The author also generalized the results of \cite{BonyHafner2008} to rotating black holes and obtained the resonance expansions. 

In \cite{Hintz2015} Hintz provides asymptotics, decay and resonance expansions for tensor-valued waves on perturbations of Schwarzschild-de Sitter spaces in all spacetime dimensions $n\geq 4.$

We mention also the recent works \cite{Gannot2014} and \cite{Warnick2015}, where quasi-normal modes in rather different geometry of Anti-de-Sitter black holes are discussed.

 Here we consider scattering of massless uncharged Dirac fields propagating in the outer region of  de Sitter-Reissner-Nordstr{\"o}m black hole, which is spherically symmetric charged exact solution of the Einstein-Maxwell equations.  We refer to \cite{DaudeNicoleau2011} for detailed study in this background including complete time-dependent scattering theory.

The considered massless Dirac fields are represented by 2-components spinors $\psi$ belonging to the Hilbert space $L^2(\R\times \S^2;\,\C^2)$ which satisfy the evolution equation
\[\lb{1.5}i\partial_t\psi=\left(\sigma_3D_x+\alpha(x)D_{\S^2}\right)\psi\] where $\sigma_3={\rm diag}(1,-1),$ $D_x=-i\partial_x$ and $D_{\S^2}$ denotes the Dirac operator on the $2D-$sphere $\S^2.$ The potential $\alpha$ is defined in  (\ref{a2}) and contains all the information of the metric through the function $F.$ Moreover, $\alpha(x)$ decreases exponentially at both infinities:\begin{align} &\exists\,\,\alpha_\pm >0,\,\,\pm\kappa_\pm <0\qq\mbox{such that} \label{assa} \\ &\alpha(x)=\alpha_\pm e^{\kappa_\pm x}+{\mathcal O}\left( e^{3\kappa_\pm x}\right),\qq
\alpha'(x)=\alpha_\pm \kappa_\pm e^{\kappa_\pm x}+{\mathcal O}\left( e^{3\kappa_\pm x}\right)\qq\mbox{as}\,\,x\rightarrow\pm\infty.\nonumber
\end{align} 

\vspace{0.3cm}

\no{\bf Dirac on SSAH manifolds.} Note that Dirac operator  \[\lb{defD}\cD=\sigma_3D_x+\alpha(x)D_{\S^2}\]  in the exterior region of  \dSRN black hole takes the same form as a representation of Dirac operator $\cD_\sigma$ on the so called Spherically Symmetric Asymptotically Hyperbolic (SSAH) Manifolds  $\Sigma=\R_x\times\S^2_{\theta,\varphi}$ (see \cite{Daudeetal2013}) equipped with the Riemannian metric
\[\lb{falpha} \s=dx^2+\alpha^{-2}(x)d\omega^2,\] where $d\omega=d\theta^2+\sin^2\theta d\varphi^2$ is the Euclidean metric on  $\S^2.$  The assumptions on the function  $\alpha(x)$ - that determines completely the metric - are (\ref{assa}) with $\alpha\in C^2(\R),$ $\alpha >0.$

 Under these assumptions, $(\Sigma,\sigma)$ is clearly a spherically symmetric Riemannian manifold with two asymptotically hyperbolic ends $\{x=\pm\infty\}$ and the metric $\sigma$ is asymptotically a small perturbation of the ``hyperbolic like'' metrics
$$ \sigma_\pm=dx^2+e^{-2\kappa_\pm}d\omega_\pm^2,\qq x\rightarrow\pm\infty,$$ where $d\omega_\pm^2=1/(\a_\pm^2)d\omega^2$ are fixed metrics on $\S^2.$ Hence, the sectional curvature of $\sigma$ tends to the constant negative values $-\kappa_\pm^2$ on the corresponding ends $\{x\rightarrow\pm\infty\}.$ 

Such spherically symmetric manifolds are very particular cases of the much broader class of asymptotically hyperbolic manifolds (see references in \cite{Daudeetal2013}). We mention also \cite{Vasy2013} for a very general analysis of meromorphic continuation for de Sitter black holes and perturbations.

The analytically extended resolvent of Dirac operator $\cD$ on asymptotically hyperbolic manifolds was described in  \cite{Guillarmouetal2010} using the  parametrix construction extending the ideas from \cite{MazzeoMelrose1987} and  \cite{Guillarmou2005}. 

The massless Dirac operator on $(\Sigma,\sigma)$ $\cD_\sigma=\sigma_3D_x+\alpha(x)D_{\S^2}$  is self-adjoint on the Hilbert space $\cH=L^2(\Sigma;\C^2)$ and has absolutely continuous spectrum. Thus one can define its resolvent   in two ways
$$\cR_+(i\epsilon):=(\cD_\sigma -i\epsilon)^{-1},\qq \cR_-(i\epsilon):=(\cD_\sigma +i\epsilon)^{-1},\qq \epsilon >0,$$ as analytic families of bounded operators on $\cH.$

 From  \cite{Guillarmouetal2010}, Theorem 1.1, it follows that the resolvents $$\cR_\pm(\l):\qq C_0^\infty (\Sigma;\C^2)\,\,\mapsto\,\,C^\infty (\Sigma;\C^2)$$ have meromorphic continuation to $\l\in\C$ with isolated poles of finite rank.

These properties  can be transmitted  to the operator $ \cD$ using its identification with a representation of   $\cD_\sigma$ as in \cite{Daudeetal2013}.  Dirac operator $\cD$ is self-adjoint on $\cH:=L^2(\R\times\S^2, dxd\omega;\C^2),$  its spectrum is purely absolutely continuous and is given by $\R.$ 

The Riemann surface of the resolvent of the Dirac operator $\cR(\l):=(\cD-\l)^{-1}$ consists of two disconnected sheets $\C.$ We will adopt a convention that $\cR(\l)$ is defined on $\C_+$ (which corresponds to the choice of $\cR_+$ above). 
The {\em resonances} are the poles in $\C_-$ of  a meromorphic continuation  of  the cut-off resolvent  
$$\cR_\chi(\l)=\chi(\cD-\l)^{-1}\chi,\qq \chi\in C_0^\infty(\R;\C^2),$$ from the upper half-plane to $\C.$

Note that equivalently we can consider the resolvent on the  lower half plane $\C_-$ and obtain a meromorphic continuation to $\C_+$ (which corresponds to the choice of $\cR_-$ above).

\vspace{0.3cm}

Note that $\l=0$ is not resonance for Dirac operator $\cD$ as was explicitly shown in \cite{DaudeNicoleau2011}, Remark 3.7.

We consider  the scattering of massless uncharged Dirac waves towards the two ends 
 $\{x\rightarrow\pm\infty\}$ in the context of \dSRN black holes.

Note that the situation is similar to the scattering problem for the wave equation on \dSS metric. The scattering phenomena there (see  \cite{SaBarretoZworski1997}) are governed by the Schr\"odinger operator \[\lb{defP}\cP^{\rm dSS}=D_x^2+\alpha^2[\Delta_\omega+2\alpha \alpha' r^3+2\alpha^2 r^2]\] as operator in $(x,\omega)$ on $L^2(\R\times \S^2;\,\C),$ where $\alpha$ is as in (\ref{a2}) but with $Q=0$ and $r=r(x)$ via Regge-Wheeler transformation (\ref{Regge}). Here $\Delta_\omega$ is the (positive) Laplacian on $\S^2.$ The resonances for \dSS black holes are defined as the poles $\l\in \C_-$ of the meromorphic continuation of the cut-off resolvent
$$\cR_\chi^{\rm dSS}(\l)=\chi(\cP^{\rm dSS}-\l^2)^{-1}\chi,\qq \chi\in C_0^\infty(\R),$$ from $\C_+$ to $\C.$

The resonances are approximated by the lattice associated to the trapped set which is a sphere of partially hyperbolic orbits - {\em photon sphere} (see \cite{Gerard1988}, \cite{GerardSjostrand1987}). Due to radial symmetry, after separation of variables and a Regge-Wheeler transformation the problem is reduced to a family of one-dimensional Schr\"odinger operators on a line with potentials exponentially decaying at infinity and having unique non-degenerate maxima. Using the inverse of the angular momentum as a semi-classical parameter, the result of  \cite{Sjostrand1986} gives the leading order in the expansion of resonances (see  \cite{SaBarretoZworski1997}). 

In \cite{Iantchenko2015} we show that resonances for \dSRN black holes can be obtained as solutions of one-dimensional Schr\"odinger equations with similar properties as in \dSS case. Moreover, using the method of semi-classical  Birkhoff normal form (as in \cite{Iantchenko2007}, \cite{Iantchenko2008}) we obtain complete asymptotic 
expansions in both \dSS and \dSRN cases.

From the physicists point of view, the quasi-normal modes for Reissner-Nordstr{\"o}m black holes were calculated numerically in \cite{WuZhao2004} (massless case), \cite{ChangShen2007} (massive case) and \cite{Jing2004} (de Sitter variant of the massless case). 
Note that the authors treated the Dirac resonances exactly as solutions of the Schr{\"o}dinger equation similar to (\ref{Schr}) (see also \cite{Chandrasekhar1983}, \cite{Chandrasekhar1980}). In  \cite{Iantchenko2015} Theorem 1 shows a different  point of view and gives exact relation between Schr{\"o}dinger and Dirac resonances.  Indeed, due to the symmetry of the equation, the set of  non-zero Schr{\"o}dinger resonances consists of two sets interposed:
the set of Dirac resonances and its mirror image  with respect to the imaginary axis.

Our reason to consider massless and uncharged fields is that the resulting Dirac operator 
coincides with a representation of a $\cD$ on the Spherically Symmetric Asymptotically Hyperbolic Manifolds $\Sigma$ as above and the global properties of its resolvent are already known thanks to  \cite{Guillarmouetal2010}. Moreover, the one-dimensional massless Dirac operator is a 2-by-2 matrix operator with exponentially decreasing potential, whereas in the massive charged case it must be a 4-by-4 matrix operator with the potential decreasing exponentially to some non-zero constants at infinities (see \cite{DaudeNicoleau2010} and \cite{Gobin2015}). For the massless uncharged fields the Dirac operator 
has supersymmetric structure (see  \cite{Chandrasekhar1980}, \cite{Thaller1992} and \cite{Jing2004}) and
 has a nice relation to a Schr{\"o}dinger operator similar 
to that appearing  in scattering problem for the wave equation in \dSS metric (see  \cite{SaBarretoZworski1997}).

 The formulas obtained in this paper for the massless uncharged case indicate what one should expect to get in the general case as it is believed that, due to intense gravitation near the event and cosmological horizons of the black hole, even if the Dirac fields are massive, they propagate asymptotically as in the massless case (see \cite{Gobin2015}, \cite{Iantchenko2016}).

\vspace{5mm}

{\bf Resonance expansions.}
Our main result concerns an expansion of the solution of the evolution equation for the massless Dirac fields (\ref{1.5}) in terms of resonances, and consider the decay of the solution. We obtain the similar formulas as in  \cite {BonyHafner2008} with the principal difference due to the fact that $\l=0$ is not a resonance in the massless Dirac case. 


Let $\psi_0\in \cH=L^2(\R\times \S^2;\C^2).$ Then there exists a unique solution $\psi(t)\in C(\R_t;\cH)$ satisfying
$$\left\{\begin{array}{rl} i\partial_t\psi(t)&=\cD \psi(t)\\
\psi(0)&=\psi_0,\end{array}\right.$$ and this solution is given by $\psi(t)=e^{-it\cD}\psi_0.$ Moreover, the energy is conserved along the evolution
\[\lb{consen}\| e^{-it\cD}\psi_0\|_\cH= \|\psi_0\|_\cH.\]

Now, we pass to formulation of our main result.

For $\chi\in C_0^\infty(\R)$  we denote the cut-off resolvent  
$ \cR_\chi(\l)=\chi(\cD -\l)^{-1}\chi $ as before.
For a resonance $\l_j$ we denote $m(\l_j)$ the multiplicity of $\l_j.$ Then we have a Laurent expansion of the cut-off resolvent near $\l_j:$
$$  \cR_\chi(\l)=\sum_{k=1}^{m(\l_j)}\frac{A_k}{(\l-\l_j)^k}+A(\l,\l_j),$$ where $A(\cdot,\l_j)$ is holomorphic near $\l=\l_j.$

We define $\pi_{j,k}^\chi$ by \[\lb{1.10} \pi_{j,k}^\chi=-\frac{1}{2\pi i}\oint\frac{(-i)^k}{k!}\cR_\chi(\l)(\l-\l_j)^kd\l.
\]

Our main result is the following

\begin{theorem}\lb{th-1.3} Let $\chi\in C_0^\infty(\R)$ and $t$ be large enough. Let $ \cD=\sigma_3D_x-\alpha(x)D_{\S^2}$ be the Dirac operator (\ref{defD}). \\
\no (i) We choose $\mu >0$  such that there is no resonance $\l\in{\rm Res}\,(\cD)$ with $\Im\l=-\mu$ and $\mu\not\in\{ \Im\mu_{k,l}^r\},$ where $\mu_{k,l}^r$ are pseudopoles, defined later.  Then there exists $M>0$ with the following properties.

 Let $u\in \cH= L^2(\R\times \S^2;\,\C^2)$ such that $\langle D_{\S^2}\rangle^M u\in \cH .$ 
\[\lb{1.11}
\chi e^{-it\cD}\chi u= \!\!\!\!\sum_{\small\begin{array}{l}\l_j\in {\rm Res}\,(\cD)\\ \Im\l_j >-\mu\end{array}}\!\!\sum_{k=0}^{m(\l_j)-1}e^{-i\l_j t}t^k\pi_{j,k}^\chi u+E_1(t)u
\] with \[\lb{1.12}\|E_1(t)u\|_{\cH} \lesssim e^{-\mu t}\|\langle D_{\S^2}\rangle^Mu \|_{\cH},  \]
and the sum is absolutely convergent in the sense that 
\[\lb{1.13}\sum_{\small\begin{array}{l}\l_j\in {\rm Res}\,(\cD)\\ \Im\l_j >-\mu\end{array}}\!\!\sum_{k=1}^{m(\l_j)-1}\|\pi_{j,k}^\chi \langle D_{\S^2}\rangle^{-M}\|_{\cL(\cH,\cH)}\lesssim 1.\]

\no (ii) Suppose $g\in C([0,+\infty )),$ $\lim_{|x|\rightarrow\infty}= 0,$ is positive, strictly decreasing function with $x^{-1}\leq g(x)$ for $x$ large. Let $u\in \cH$ be such that $\left(g(D_{\S^2})\right)^{-1}u\in \cH.$  

Then there exists $\epsilon >0$ such that
\[\lb{1.14}
 \|\chi e^{-it\cD}\chi u\|_{\cH}\lesssim g(e^{\epsilon t})\|\left( g(D_{\S^2})\right)^{-1}u\|_{\cH}.\]

\end{theorem}
{\bf Remark 1.} By part (ii) of the theorem for $u\in\cH$ the local energy is integrable if $(\ln\langle D_{\S^2}\rangle )^\alpha u\in\cH $ for some $\alpha >1$ (by choosing $g(x)=(\ln x)^{-\alpha}$), and decays exponentially if $\langle D_{\S^2}\rangle^\epsilon u\in\cH$ for some $\epsilon \in (0,1)$ (by choosing $g(x)=x^{-\epsilon}$). Choose $\alpha >1,$ then we get

\[\lb{1.18}
 \|\int_0^\infty\chi e^{-it\cD}\chi udt\|_{\cH}\lesssim \|\left( \ln\langle D_{\S^2}\rangle \right)^\alpha u\|_{\cH}.\]

{\bf Remark 2.} The results of Theorem \ref{th-1.3} extend to Dirac operators $\cD=\cD_\sigma$ on spherically symmetric asymptotically hyperbolic (SSAH) manifolds  $(\Sigma,\sigma)$ as in the Introduction supposing that \\
\no i) function $\alpha>0$  in (\ref{falpha}) is analytic satisfying (\ref{assa}) and extends as to a holomorphic function in a conic \neigh{} of the real axis given by $|\arg w|<\theta$ and $\alpha$ satisfies there
$$|\alpha(w)|\leq C\exp (-|w|/C),\qq \Re w\rightarrow\pm\infty; $$

\no ii) $\alpha$ has a unique non-degenerate maximum at some point $x_0.$

\vspace{3mm}

{\bf At last I will comment on our choice of the model.}
Our reason to study massless and uncharged fields is that the resulting Dirac operator 
coincides with a representation of a $\cD$ on the Spherically Symmetric Asymptotically Hyperbolic Manifolds $\Sigma$ as above and the global properties of its resolvent are already known due to  \cite{Guillarmouetal2010}. Moreover, the one-dimensional massless Dirac operator is 2-by-2 matrix operator and
has the most simple yet nontrivial structure and has a nice relation to a Schr{\"o}dinger operator similar 
to that appearing  in scattering problem for the wave equation in \dSS metric (see  \cite{SaBarretoZworski1997}). As the last problem is well studied, we can easily transmit  many already existing results to the Dirac case, and  apply the Birkhoff normal form construction. This paper is the second one in our project on quasi-normal modes for Dirac fields in black holes geometry  and many properties and methods from this model can be generalized to more complicated situations as the cases of massive charged Dirac fields and rotating (Kerr-Newman) black holes (see \cite{Iantchenko2016} and forthcoming papers).

\vspace{3mm}

{\bf The paper is organized as follows.}\\
 In Section \ref{Sect2} we collect the main properties of the massless Dirac operator on \dSRN metric using decomposition of the Hilbert space $\cH$ in spin-weighted harmonics. We recall Regge-Wheeler change of variables and describe the properties of the potential $\alpha(x).$ Moreover, we recall our results from \cite{Iantchenko2015} on localization of quasi-normal frequencies.\\
In Section \ref{s-resest} we focus on cut-off resolvent estimates crucial for the proof of Theorem \ref{th-1.3}. Subsection \ref{ss-rel} provides with general relations between the cut-off Dirac and  the cut-off Schr{\"o}dinger resolvents. In Subsection \ref{s-mainres} we formulate and prove Theorem \ref{th-2.1} containing the estimates on the cut-off resolvents in different
zones on the complex plane.\\
In Section \ref{s-Resexp} we prove Theorem \ref{th-1.3}.

\section{Preliminaries.}\lb{Sect2}
 In this section we summarize the properties of  the Dirac operator  and formulate the main results. 

By decomposition (see \cite{DaudeNicoleau2011}) of the Hilbert space $\cH=L^2(\R\times\S^2, dxd\omega;\C^2)$ in spin-weighted spherical harmonics $F_m^l,$ $(l,m)\in\cI,$
\[\lb{dec1}\cI=\{(l,m);\,\,l-\frac12\in\N,\,\,l-|m|\in\N\},\qq\cH=\bigoplus_{(l,m)\in\cI}\cH_{l,m},\qq  D_{\S^2}F_m^l=-(l+\frac12)\sigma_1 F_m^l,\]
where $\cH_{l,m}$ is identified with $L^2(\R;\C^2),$
 we obtain the orthogonal decomposition for the Dirac Hamiltonian $\cD$
\[\lb{dec2}\cD=\bigoplus_{(l,m)\in\cI}\cD^{l,m},\qq \cD^{l,m}:=\cD_{|\cH_{l,m}}=\s_3D_x-\left(l+\frac12\right) \a(x)\s_1,\] where
the one-dimensional Dirac operator $\cD^{l,m}$ does not depend on index $m.$ 

Now, the scattering of massless charged Dirac fields in de Sitter-Reissner-Nordstr{\"o}m black holes is described (see \cite{DaudeNicoleau2011}) by the scattering on the line for the massless Dirac system
\begin{equation}\label{DiracSystem}
\left[ \s_3D_x-n \a(x)\s_1\right]\psi=\l\psi,\qq n=l+\frac12\in\N,\qq \s_3=\left(
         \begin{array}{cc}
           1 & 0 \\
           0 & -1 \\
         \end{array}
       \right)\qq \s_1=\left(
         \begin{array}{cc}
           0 & 1 \\
           1 & 0 \\
         \end{array}
       \right),
\end{equation}
which is a special form of Zakharov-Shabat system (see \cite{IantchenkoKorotyaev2013} with $q=-n \a(x)\in\R$).  The potential $\a(x)$ is given by
\begin{equation}\label{a2}
\a^2(x)=\frac{F(r(x))}{r^2(x)},\qq F(r)=1-\frac{2M}{r}+\frac{Q^2}{r^2}-\frac{\Lambda}{3}r^2,
\end{equation}
where $M>0,$ $Q\in\R$ are the mass and the electric charge of the black hole respectively, $\L >0$ is the cosmological constant.
The equation (\ref{DiracSystem}) is expressed by means of Regge-Wheeler coordinate $x$ related to the original radial coordinate $r$ by means of the equation
\[\lb{Regge}\frac{dx}{dr}=\frac{1}{F(r)}.\]
 We suppose that $Q^2 <\frac98M^2$ and $\L M^2$ is small enough. Then the function $F(r)$ has four real zeros
 $$r_n < 0< r_c<r_-<r_+.$$ The sphere $\{r=r_c\}$ is called the Cauchy horizon, whereas the spheres $\{r=r_-\}$ and $\{r=r_+\}$ are the event and cosmological horizons respectively.

  We consider scattering in the exterior region $\{ r_- <r<r_+\},$ where we have \[\lb{expdec}  \a(x)\sim \a_\pm e^{\kappa_\pm x}\qq\mbox{as}\,\,x\rightarrow\pm\infty\,\,\mbox{or}\,\,r\rightarrow r_\pm,\] where
 $\kappa_- >0,$  $\kappa_+<0$ are surface gravities at  event and cosmological horizons respectively,  $\alpha_\pm$ are fixed constants depending on the parameters of the black hole.

It is well known (see \cite{DEGM})   that the operator $\s_3D_x-n \a(x)\s_1$ acting in $L^2(\R )\os L^2(\R )$ is
self-adjoint and its spectrum  is purely absolutely
continuous and is given by the set $\R$. In \cite{IantchenkoKorotyaev2013} we studied resonances of such operators in the case of compactly supported potential $-n\a(x).$ Then the outgoing  solutions (Jost solutions) have analytic continuation from the upper half-plane $\C_+$ to the whole complex plane $\C$ and resonances are the zeros in $\C_-$ of the Wronskian for the Jost solutions or, equivalently, the poles in    $\C_-$ of the analytic continuation of the cut-off resolvent. For non-compactly supported exponentially decreasing potential $-n \a(x)$ satisfying (\ref{expdec}) such method of analytic continuation  is possible in a strip $\{\l\in\C;\,\,\Im\,\l>-\epsilon\}$ for some $\epsilon >0$ (see \cite{Froese1997}). In order to calculate resonances in a larger domain (a sector) one uses  the method of complex scaling. It is well-known that different definitions give rise to the same set of resonances in the domains where both definitions are applicable (see \cite{HelfferMartinez1987}).

We use that  (\ref{DiracSystem}) can be written in the semi-classical way as
$$
\cD_{-\alpha}\psi\equiv\left[ h\s_3D_x-\alpha(x)\s_1\right]\psi=z \psi,\,\, z=\l/n,$$ 
with ``Planck constant'' $ h=1/n.$ 
We denote the set of resonances for $\cD_{-\alpha}=h\s_3D_x-\alpha(x)\s_1$ by ${\rm Res}\,(\cD_{-\alpha})\subset \C_-.$
Note the following symmetry property of the Dirac operator $\cD_{-\alpha}$ with real-valued $\alpha:$  $$\l\in{\rm Res}\,(\cD_{-\alpha})\,\,\Leftrightarrow\,\,-\overline{\l}\in{\rm Res}\,(\cD_{\alpha}).$$

We consider also the semi-classical Schr\"odinger operator  \[\lb{Schr}P=h^2(D_x)^2+V_h(x),\qq V_h(x)=\a^2(x)+h\a'(x).\] We say that $\l\in\C_-$ is a resonance for $P$ if for some function $\chi\in C_0^\infty(\R)$ $\l$ is a a pole  of meromorphic continuation of the cut-off resolvent
$\chi(P-\l^2)^{-1}\chi.$
We denote the set of resonances of $P$ by ${\rm Res}\,(P).$ The set of resonances is invariant under the change of sign $\alpha$ $\mapsto$ $-\alpha$ and invariant under  the reflection ${\rm S}$ with respect to $i\R:$
$\l\in{\rm Res}\,(P)\,\,\Leftrightarrow\,\,-\overline{\l}\in{\rm Res}\,(P).$

For a  set of points $\sigma=\{\l_j\}\in\C_-$ we denote the mirror image with respect to $i\R$ by \[\lb{mirror}\sigma^{\rm S}:=\{-\overline{\l_j}\}\in\C_-.\]

In \cite{Iantchenko2015} we show that the following relation between resonances for $\cD_{\pm \alpha}$ and $P$ :
\[\lb{L-UnD}{\rm Res}\,(P)\setminus\{0\}={\rm Res}\,(\cD_{-\alpha})\cup{\rm Res}\,(\cD_{\alpha})={\rm Res}\,(\cD_{-\alpha})\cup{\rm Res}^{\rm S}\,(\cD_{-\alpha}).\]

The principal symbol of the potential in (\ref{Schr}) $V_0(x)=\a^2(x)$ has a non-degenerate maximum at $x_0=x(r_0),$ where $ r_0=3M/2+\sqrt{(3M)^2/4-2Q^2}$ and
$$
V_0(x_0)=r_0^{-4}\left(Mr_0-Q^2-\frac{\L}{3}r_0^4,\right),\qq V_0''(x_0)=-2\left(\frac{3M}{r_0}-\frac{4Q^2}{r_0^2}\right)V_0^2(x_0).$$ It is well-known \cite{Sjostrand1986} that the resonances associated to the non-degenerate maximum of the principal symbol $V_0(x)$ of potential, barrier top resonances, are close to the lattice of pseudopoles.

Note that 
resonances (quasi-normal modes) for an operator similar to (\ref{Schr})    were  mathematically studied in \cite{SaBarretoZworski1997} and \cite {BonyHafner2008} in  the context of \dSS black holes. The authors of \cite{SaBarretoZworski1997} give two leading terms in the asymptotic expansions of resonances. In \cite{Iantchenko2015} we show that similar results also hold for the \dSRN resonances. Namely, we show that in semi-classical limit $h=1/(l+1/2)\rightarrow 0$ the resonances are close to the lattice of pseudopoles. Moreover, using  the method of  semi-classical (or quantum)  Birkhoff normal form (abbreviated qBnf, see \cite{KaidiKerdelhue2000} and \cite{Iantchenko2008}) we get the complete asymptotic expansions for the resonances both in \dSRN and \dSS cases. 

Now, using  the explicit reconstruction procedure of the qBnf as in  \cite{CdVerdiereGuillemin2011} we get explicit formulas for the next (third) order terms in the expansions of resonances.

We recall the main result of  \cite{Iantchenko2015}, Theorem 1.

{\em Let $$\Omega_{C}=\{\l\in\C_-;\,\,\Im\l>-C,\,\,\Re\l >K,\,\,\Im\l >-\theta |\Re\l| \}.$$ 
Fix a number  $N\in\N.$ Then there exist   $K>0,$  $\theta>0,$  $r\in\N$ and functions $f_j=f_j(2k+1)={\mathcal O}((2k+1)^j,$ $k\geq 0,$ $j=1,\ldots,r,$ polynomial in $2k+1$ of order $\leq j,$ such  that
 for any $C>0$  there exists an injective map, $b_N$ from the set of pseudo-poles
\begin{align*}\mu_{k,l}^r=&  (l+1/2)\left(z_0+\frac{f_1(2k+1)}{l+1/2}+\frac{f_2(2k+1)}{(l+1/2)^2}+\ldots+\frac{f_{r+1}(2k+1)}{(l+1/2)^{r+1}}\right),\,\,l\in\N,\,\,k\in\N_0,
\end{align*}
into the set of resonances \[\lb{UnD}{\rm Res}\,(\cD)\cup{\rm Res}^{\rm S}\,(\cD),\qq \cD=\sigma_3D_x-\alpha(x)D_{\S^2},\]  such that all the resonances in $\Omega_{C}$ are in the image of $b_N$ and for $b_N(\l)\in\Omega_{C},$
$$b_N(\l)-\l={\mathcal O}\left(|\l |^{-N}\right). $$ Here  $\{\cdot\}^{\rm S}$ denotes the mirror reflection of the set $\{\cdot\}\in\C_-$ in $i\R$ (see (\ref{mirror})) and \begin{align*}&z_0=\alpha(x_0),\,\,\omega=\left(\frac12|V_0''(x_0)|\right)^\frac12,\\
&f_1=- \frac{i\omega}{2 z_0}  (2k+1),\,\, f_2=-\frac{i\omega}{2 z_0}  (2k+1)\left[-\frac{1}{4iz_0^2}\omega(2k+1)+\frac{1}{2i} b_{0,2}(2k+1)+b_{1,2} \right],\\ &b_{0,2}=\frac{15}{4\cdot 12^2}\frac{(V_0'''(x_0))^2}{\o^5}+\frac{V_0''''(x_0)}{32\o^3},\qq b_{1,2}=\frac{1}{8z_0^3}-\frac{3}{8z_0\o^2}V_0'''(x_0).
 \end{align*}

 The resonance in ${\rm Res}\,(\cD)$ corresponding to pseudopole $\mu_{k,l}^r$ has multiplicity $2l-1.$}


\section{The resolvent estimates.}\lb{s-resest}

\subsection{Relation between  the cut-off resolvents of  1D Dirac and Schr\"odinger operators.}\lb{ss-rel}
Here we recall some well-known properties of one dimensional  massless Dirac operators mostly following 
\cite{Iantchenko2015} and  \cite{IantchenkoKorotyaev2013} .

In $\cH=L^2(\R )\os L^2(\R )$ we consider massless Dirac operator \[\lb{Diracmassles}\cD_q=\cD_0+V:=\ma -i\partial_x&0\\ 0&i\partial_x\am     +\ma 0&q\\ q^\star&0\am,\qq q\in L^1(\R;\C),\] and  the  Dirac equation $\cD_q f=\l f.$

Operator $\cD_q$ is self-adjoint in $\cH$ and its spectrum is absolutely continuous   $\sigma(\cD_q)=\R$ with no bound states.

Now, we suppose that $q$ is real-valued and  smooth.  Note that the method below  also works for $q=ir$ pure imaginary. 
For the Dirac operator
$${\cD_q}=-i\s_3\partial_x+q\s_1=  \ma -i\partial_x & q\\q & i\partial_x\am$$  we 
consider also its square $$\cD_q^2=-I_2\partial^2_x +\ma q^2&-iq'\\iq'&q^2\am,$$ which is matrix Schr{\"o}dinger operator. Operator $\cD_q^2$ is self-adjoint in  $\cH=L^2(\R )\os L^2(\R )$ and unitary equivalent to
\[\lb{diagon}U\cD_q^2U^{-1}=\ma \cP-&0\\
0&\cP^+\am,\qq \cP^\pm=-\partial_x^2 +q^2\pm q'.\]
Here, $$U=\frac{i}{\sqrt{2}} \ma 1 & i\\1 & -i\am,\qq U^{-1}=-\frac{i}{\sqrt{2}} \ma 1 & 1\\-i & i\am .$$

Note that, if in  (\ref{Diracmassles}) we had supposed that $q=ir,$ $r$ is real-valued, then,  as above, 
we could transform the $\cD_{ir}$ to the form  (\ref{diagon}) with $\cP^\pm=-\partial_x^2 +r^2\pm r'$ (and with different $U$ of cause).

The resolvents $(\cD_q-\l)^{-1},$ $(\cD_q^2-\l^2)^{-1},$ $(\cP^\pm-\l^2)^{-1}$ are analytic functions of $\l$ in $\C_+.$ 

In our context $q(x)=-n\alpha(x),$ $n=l+1/2,$ with $\alpha$ satisfying  (\ref{a2}) and (\ref{expdec}). Moreover, (see Proposition 1 in \cite{Iantchenko2015} coming originally from   \cite{SaBarretoZworski1997}) the function  $\alpha(x)$ is analytic on $\R$ and extends to a holomorphic function in a conic neighborhood of the real axis given by $|\arg w|<\theta$ and $\alpha$ satisfies there
\[\lb{bound_conic}|\alpha(w)|\leq C\exp (-|w|/C),\qq \Re w\rightarrow\pm\infty. \]

We introduce semi-classical parameter $h=1/n$ and semi-classical operators
\[\lb{semop}\cD_h=-\sigma_3hD_x-\alpha(x),\qq \cP_h^\pm=(hD_x)^2+V_h^\pm(x),\qq V_h^\pm(x)=(\a(x))^2\pm h\a'(x).\]
The frequency parameter scales as $z=\l/n.$

Then using the method of complex scaling (see  \cite{Iantchenko2015},  \cite{SaBarretoZworski1997},  \cite{Sjostrand1997})  we can define a meromorphic continuation of the resolvents.

The resonances for $\cD_h$ respectively $\cP^\pm_h$ are the poles $\l\in\C_-$ of meromorphic continuation
of $$\chi(\cD_h-z)^{-1}\chi,\qq \chi\in C_0^\infty(\R,\C^2),\qq\mbox{resp.}\qq\chi_1(\cP^\pm_h-z^2)^{-1}\chi_1,\qq \chi_1\in C_0^\infty(\R,\C).$$ 

The set of resonances is invariant under the change of sign $\alpha$ $\mapsto$ $-\alpha$ and invariant under  the reflection ${\rm S}$ with respect to $i\R:$
$\l\in{\rm Res}\,(P)\,\,\Leftrightarrow\,\,-\overline{\l}\in{\rm Res}\,(P).$
The relation between the resonance sets is given by (\ref{L-UnD}) 
$${\displaystyle {\rm Res}\,(\cP^\pm_h)\setminus\{0\}={\rm Res}\,(\cD_h)\cup{\rm Res}^{\rm S}\,(\cD_h)\in\C_-.}$$

Note the following  resolvent identity
\[\lb{resDiSch}
({\cD_h}-z)^{-1}=({\cD_h}+z)U^{-1}\ma (\cP_h^--z^2)^{-1}&0\\
0&(\cP^+_h-z^2)^{-1}\am U.
\]
Let $$H^s=\{ u\in L^2(\R,\C),\,\,\|u\|_{H^s} <\infty\},\,\, \|u\|_{H^s}^2:=\sum_{k=0}^s\int_\R|(h\partial_x)^k u(x)|^2 dx $$ be the standard semi-classical Sobolev spaces, and we define
$\cH^s=H^s\os H^s. $

We will extensively use the following two lemmas on the norms of the cut-off resolvents.
\begin{lemma}\lb{p-3.1} Let $\cD_h,$ $\cP^\pm_h$ be as in (\ref{semop}).   Let $\chi\in C_0^\infty(\R;\C^2)$ and $\chi_i\in C_0^\infty (\R;\C),$ $i=1,2.$ Then
\[\lb{3.1} \|\chi({\cD_h}-z)^{-1}\chi\|_{\cL\left(\cH^0,\cH^{0} \right)}\lesssim\langle z\rangle\left(\|\chi_1(\cP_h^--z^2)^{-1}\chi_1\|_{\cL\left(H^0,H^{1} \right)}+\|\chi_2(\cP^+_h-z^2)^{-1}\chi_2\|_{\cL\left(H^0,H^{1} \right)}\right).\]
\end{lemma}
{\bf Proof.}  We have $$\chi({\cD_h}- z)^{-1}\chi=\chi({\cD_h}+z)(\cD_h^2-z^2)^{-1}\chi=
({\cD_h}+z)\chi(\cD_h^2-z^2)^{-1}\chi-[{\cD_h},\chi](\cD_h^2-z^2)^{-1}\chi.
$$ For $u\in C^\infty$ we have
$$\|( h\s_3D_x-\alpha(x)\s_1+z)\chi u\|_{\cH^0}\lesssim\langle z\rangle\|\chi u\|_{\cH^1},$$
where we used that $\alpha\in L^\infty\cap L^1.$ 
Then $$\|\chi({\cD_h}-z)^{-1}\chi u\|_{\cH^{0} }\lesssim\langle z\rangle\|\chi (\cD_h^2-z^2)^{-1}\chi u\|_{\cH^1}$$ which implies (\ref{3.1}) due to (\ref{diagon}).
\hfill\BBox

\begin{lemma}\lb{l-interpol} Let $\cP_h$ be either $\cP_h^+$ or $\cP_h^-$ defined in (\ref{semop}) and   $\chi\in C_0^\infty (\R;\C).$ Then for $j=1,2$
\[\lb{interpol} \|\chi(\cP_h-z^2)^{-1}\chi\|_{\cL\left(H^0,H^{j} \right)}\lesssim\langle z\rangle^j  \|\chi(\cP_h-z^2)^{-1}\chi\|_{\cL\left(H^0,H^0 \right)}+\langle z\rangle^{j-2}\|\chi\|_{\cL\left(H^0,H^0 \right)}
\]
\end{lemma}
{\bf Proof.} We use elliptic regularity (see e.g. \cite{Zworski2012}, Theorem 7.1): if $U\subset\subset W$ then
$$\| u\|_{H^2(U)}\leq C\left(\|u\|_{L^2(W)}+\|\cP_h u\|_{L^2(W)}\right).$$ Hence, if $\tilde{\chi}\in C_0^\infty(\R)$ satisfies $\tilde{\chi}\chi=\chi$ then
\[\lb{2.2.4} \|\chi u\|_{H^2(\R)}\leq C\left(\|\tilde{\chi}u\|_{L^2(\R)}+\|\tilde{\chi}\cP_h u\|_{L^2(\R)}\right).
\] Now, we apply $(\ref{2.2.4})$ to $u=(\cP_h -z^2)^{-1}\chi f,$ $f\in L^2,$ so that
$$\|\chi (\cP_h - z^2)^{-1}\chi f\|_{H^2(\R)}\leq C\left(\|\tilde{\chi}(\cP_h -z^2)^{-1}\chi f\|_{L^2(\R)}+\|\tilde{\chi}\cP_h(\cP_h -z^2)^{-1}\chi f\|_{L^2(\R)}\right).$$ Since
$$\tilde{\chi}\cP_h(\cP_h -z^2)^{-1}\chi f=\chi f+z^2\tilde{\chi}(\cP_h -z^2)^{-1}\chi f$$
we get $$\|\chi (\cP_h - z^2)^{-1}\chi f\|_{H^2(\R)}\leq C\left((1+\langle z\rangle^2)\|\tilde{\chi}(\cP_h -z^2)^{-1}\chi f\|_{L^2(\R)}+\|\chi f\|_{L^2(\R)}\right)$$ and (\ref{interpol}) for $j=2$ follows.

 Now, by interpolating between $H^0$ and $H^2$ norms we get
$$\|\chi (\cP_h - z^2)^{-1}\chi f\|_{H^1(\R)}\leq  C\left(\langle z\rangle\|\tilde{\chi}(\cP_h -z^2)^{-1}\chi f\|_{L^2(\R)}+\langle z\rangle^{-1}\|\chi f\|_{L^2(\R)}\right).$$
\hfill\BBox

\subsection{Estimates on the   cut-off resolvent.}\lb{s-mainres}

The  following result is an analogue of  Theorem 2.1 in \cite{BonyHafner2008} for the \dSRN case and follows from the general result in \cite{Guillarmouetal2010}, the proof of Theorem 2.1 in  \cite{BonyHafner2008} and  Lemma \ref{p-3.1} in the previous section.

\begin{figure}[htbp]
\includegraphics[width =16 cm]{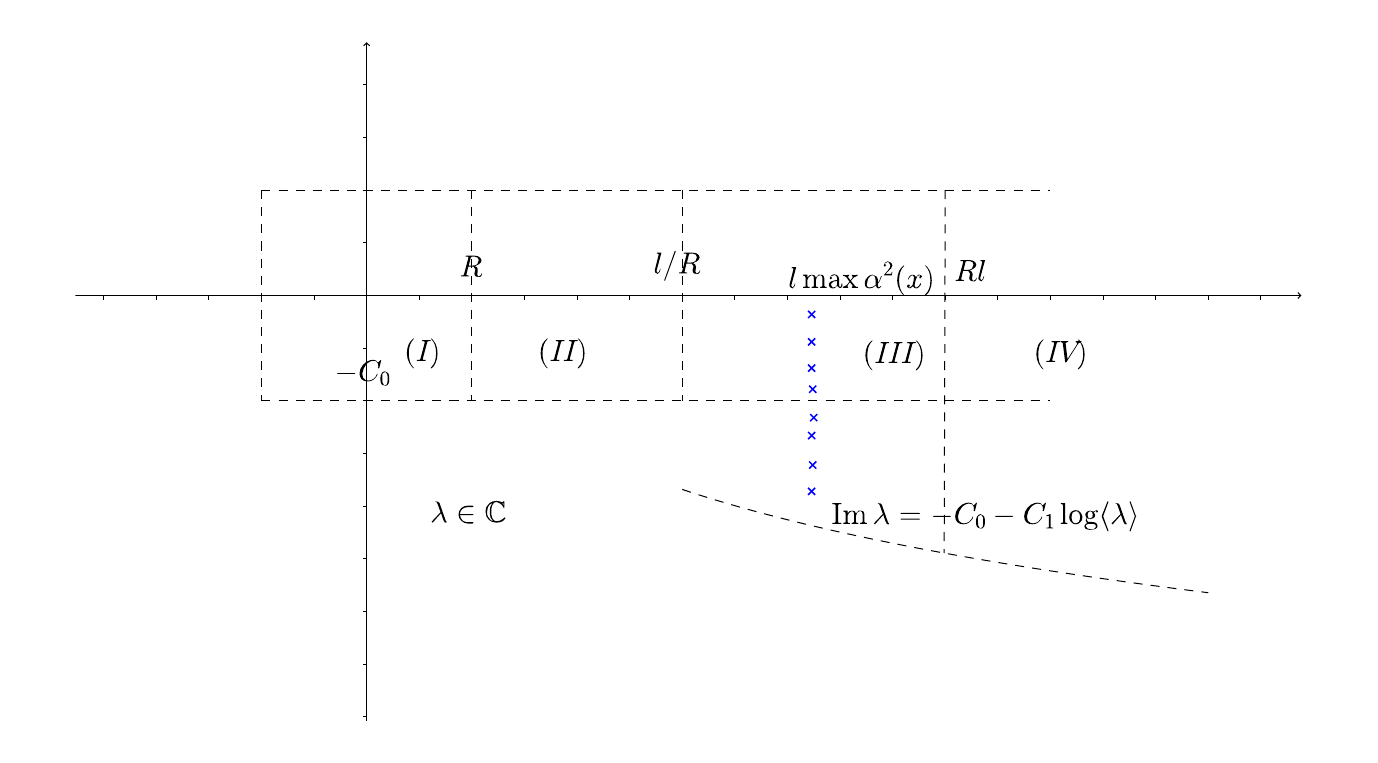}
 \caption{The different zones in Theorem \ref{th-2.1} following \cite{BonyHafner2008}. \label{Fig1} }
\end{figure}

 We obtain  bounds for the cut-off resolvents of $\cD$ defined in (\ref{defD}), the radial one-dimensional massless Dirac operator $\cD=\cD_{-n\alpha}$  and the Schr{\"o}dinger operators $\cP^\pm$  in different zones as in Fig.\,2 in \cite{BonyHafner2008} which we reproduce here as Figure \ref{Fig1}.

\begin{theorem}\lb{th-2.1} Let $\chi\in C_0^\infty(\R),$  $\cD= \sigma_3D_x-\alpha(x)D_{\S^2}$  the Dirac operator in (\ref{defD}), ${\cD_l}=-i\s_3\partial_x-\left(l+\frac12\right)\alpha\s_1$ and $\cP_l$ be either $\cP_l^-$ or $\cP_l^+,$ \[\lb{cppm}\cP_l^\pm:=-\partial^2_x+\left(l+\frac12\right)^2\alpha^2(x)\pm\left(l+\frac12\right)\alpha'(x).\] Let ${\rm Res}(\bullet)$ denote resonance sets for  either $\cP_l$ or ${\cD_l}$ counted with their multiplicities. Then, the following facts hold true uniformly in $l$:\\
\no (I) For all $R>0,$ the number of resonances of $\cP_l$ and ${\cD_l}$ is bounded in $B(0,R)$. Moreover, there exists $C>0$ such that 
\[\lb{2.2}  \|\chi({\cD_l}-\l)^{-1}\chi\|_{\cL\left(\cH^0,\cH^{0} \right)}\leq  \|\chi({\cD}-\l)^{-1}\chi\|\leq C \prod_{\small\begin{array}{l}\l_j\in {\rm Res}\,(\cD)\\ |\l_j|<2R\end{array}} \frac{1}{|\l-\l_j|},\] for all $\l\in B(0,R).$  Moreover, $0\not\in{\rm Res}(\cD).$

\no (II)  For $R$ large enough, $\cP_l$ and ${\cD_l}$ have no resonances in $[R, l/R]+i[-C_0,0].$ Moreover, there exist constants $C_1$ and $C_2,$  such that
\[\lb{2.3}\|\chi(\cP_l-\l^2)^{-1}\chi\|_{\cL\left(H^0,H^0 \right)}\leq C_1\langle \l\rangle^{-2},\]
\[\lb{2.3bis} \|\chi({\cD_l}-\l)^{-1}\chi\|_{\cL\left(\cH^0,\cH^{0} \right)}\leq C_2\langle \l\rangle^{-1},\]
for $\l\in [R,l/R]+i[-C_0,C_0].$

\no (III) Let $C_0, R$ be fixed.
 For $l$ large enough, the resonances of $\cD_l$ and $\cP_l$ in $[l/R,Rl]+i[-C_0,0]$ are given by  $b_N(\mu_{k,l}^r)$  (see (\ref{UnD})) and, in particular, their number is bounded uniformly in $l$. Moreover, there exist $C_1$ and $C_2$ such that
\[\lb{2.4}\|\chi(\cP_l-\l^2)^{-1}\chi\|_{\cL\left(H^0,H^0 \right)}\leq C_1\langle \l\rangle^{ C_1}\prod_{\small\begin{array}{l}\l_j\in {\rm Res}\,({\cP_l})\\ |\l-\l_j|<1\end{array}} \frac{1}{|\l-\l_j|},\]

\[\lb{2.4bis} \|\chi({\cD_l}-\l)^{-1}\chi\|_{\cL\left(\cH^0,\cH^{0} \right)}\leq C_2\langle \l\rangle^{ C_2}\prod_{\small\begin{array}{l}\l_j\in {\rm Res}\,({\cP_l})\\ |\l-\l_j|<1\end{array}} \frac{1}{|\l-\l_j|},\] 
for  $\l\in [l/R,Rl]+i[-C_0,C_0].$ Furthermore, $\cP_l$ and $\cD_l$ have no resonances in $[l/R,Rl]+i[-\epsilon,0]$ for some $\epsilon >0,$ and for some $C_1$ and $C_2$ the following estimates hold true
\[\lb{2.5}\|\chi(\cP_l-\l^2)^{-1}\chi\|_{\cL\left(H^0,H^0 \right)}\leq C_1\frac{\ln{\langle \l\rangle}}{\langle \l\rangle}e^{C|\Im\l |\ln\langle \l\rangle} 
\]

\[\lb{2.5bis}\|\chi(\cD_l-\l)^{-1}\chi\|_{\cL\left(\cH^0,\cH^0 \right)}\leq C_2\ln{\langle \l\rangle}e^{C|\Im\l |\ln\langle \l\rangle} \]

\no (IV) Let $C_0, C_1 >0$ be fixed. For $R$ large enough, $\cP_l,$ $\cD_l$  have no resonances in $$\{\l\in\C;\,\,\Re\,\l >R l,\,\,\mbox{and}\,\, 0\geq\Im\,\l\geq -C_0-C_1\log\langle\l\rangle\}.$$
Moreover, there exist  $C_2$ and $C_3$ such that 
\[\lb{2.6} \|\chi(\cP_l -\l^2)^{-1}\chi\|_{\cL\left(H^0,H^0 \right)}\leq \frac{C_2}{\langle\l\rangle}e^{C_2|\Im\,\l|},
\] 

\[\lb{2.6bis}  \|\chi({\cD_l}-\l)^{-1}\chi\|_{\cL\left(\cH^0,\cH^{0} \right)}\leq C_3 e^{C_3|\Im\,\l|}.\]
for $\Re\l>Rl$ and $C_0\geq\Im\l\geq -C_0-C_1\log\langle\l\rangle.$

\end{theorem}
{\bf Proof.} 
(I)  In \cite{Guillarmouetal2010} it was shown that the cut-off resolvent $\chi(\cD-\l)^{-1}\chi$ has a meromorphic extension to $\C.$ Then the number of resonances in $B(0,R)$ is always bounded, and estimate (\ref{2.2}) follows.  Moreover, $\cD$ has no eigenvalues or resonances in $\R$  (see \cite{DEGM}).

In the rest of the proof we will use the bounds on the cut-off resolvents obtained in  \cite{BonyHafner2008} for the operator $D_x^2+\alpha^2[l(l+1)+2\alpha \alpha' r^3+2\alpha^2 r^2]$ (\dSS case, $Q=0$) which are still valid for the operator $\cP$ as in (\ref{cppm}). We will scale the operators appropriately in each zone and use the relations between the semi-classical Dirac and Schr{\"o}dinger cut-off resolvents as in Lemmas \ref{p-3.1} and \ref{l-interpol}.

(II) Let  $\l\in [R,l/R]+i[-C_0,C_0].$ Estimate (\ref{2.3}) for $j=0$ follows as  in \cite{BonyHafner2008} for a similar operator: $$\|\chi(\cP_l-\l^2)^{-1}\chi\|_{\cL\left(H^0,H^0 \right)}\leq C_0\langle \l\rangle^{-2}.$$ 
Now, we consider the scaled operator
\[\lb{2.25bb}\cP_h=-h^2\partial^2_x + \alpha^2(x)\pm h\alpha'(x),\qq (\cP_l-\l^2)^{-1}=h^2(\cP_h-z^2)^{-1},\,\,h=(l+1/2)^{-1}\] where
\[\lb{2.26bb} z=h\l \in [Rh,1/R]+i[-C_0h, C_0h].\] Then
 $$\|\chi(\cP_h-z^2)^{-1}\chi\|_{\cL\left(H^0,H^0 \right)}=h^{-2}\|\chi(\cP_l-\l^2)^{-1}\chi\|_{\cL\left(H^0,H^0 \right)}\leq C_0\langle z\rangle^{-2}.$$
By Lemma \ref{l-interpol}, estimate (\ref{interpol}), we get
$$\|\chi (\cP_h - z^2)^{-1}\chi f\|_{H^2(\R)}\leq C\| f\|_{H^0(\R)},\qq \|\chi (\cP_h - z^2)^{-1}\chi f\|_{H^1(\R)}\leq C\langle z\rangle^{-1}\| f\|_{H^0(\R)}.
$$ 
Then we apply Lemma \ref{p-3.1}, (\ref{3.1}), to $\cD_h=-\s_3hD_x-\alpha\s_1$ and get
\begin{align*}& \|\chi({\cD_h}-z)^{-1}\chi\|_{\cL\left(\cH^0,\cH^{0} \right)}\\
& \leq C\langle z\rangle\left(\|\chi_1(\cP_h^--E)^{-1}\chi_1\|_{\cL\left(H^0,H^{1} \right)}+\|\chi_2(\cP^+_h-E)^{-1}\chi_2\|_{\cL\left(H^0,H^{1} \right)}\right)\leq C.\nonumber
\end{align*}
As $z=h\l,$ we get
$$\|\chi(\cD_l-\l)^{-1}\chi\|_{\cL\left(H^0,H^0 \right)}= \| h\chi({\cD_h}-z)^{-1}\chi\|_{\cL\left(\cH^0,\cH^{0} \right)}\leq Ch$$
which implies (\ref{2.3bis}).

\begin{figure}[htbp]
\includegraphics[width =13 cm]{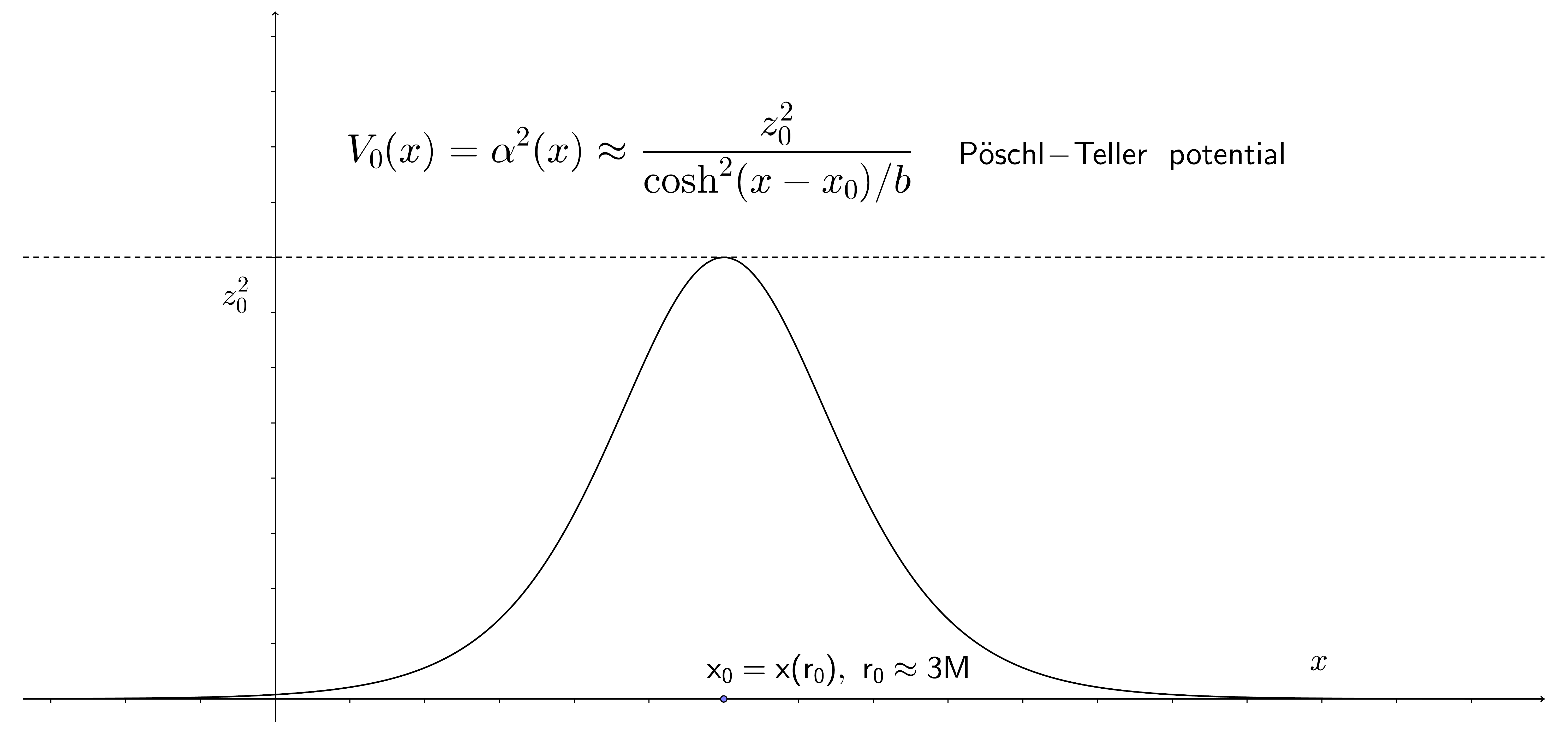}
 \caption{Example of a barrier top potential used in numerical computations.\label{Fig3} }
\end{figure}

(III) Let  $\l\in [l/R,Rl]+i[-C_0,C_0].$ Here we deal with the resonances  approximated by the lattice associated to the trapped set which is a sphere of partially hyperbolic orbits -  {\em photon sphere}.  For a fixed $l\,\,$each radial operator has a string of resonances associated to a unique non-degenerate maximum. Estimate (\ref{2.4}) follows as in \cite{BonyHafner2008} for a similar operator. As there and in (\ref{2.26bb}) in zone (II) above we define  semi-classical parameter $h=1/(l+1/2)$ and 
\[\lb{2.25}\cP_h=-h^2\partial^2_x + \alpha^2(x)\pm h\alpha'(x),\qq (\cP_l-\l^2)^{-1}=h^2(\cP_h-E)^{-1},\] where
\[\lb{2.26}E=h^2\l^2\in [R^2/2,R^2]+i[-3RC_0h, 0]\,\,\subset\,\,[a,b]+i[-ch,ch],\] for some $0<a<b$ and $0<c.$ The principle symbol of the potential $V_0=\alpha^2$ has a profile similar to the P\"oschl-Teller potential shown on Figure \ref{Fig3} and admits at $x_0$ a non-degenerate maximum at energy $E=z_0^2.$ On the other hand, for $E\neq z_0^2,$ the energy level $E$ is non trapping for $p_0(x,\xi)=\xi^2+V_0(x).$

Then the distorted operator  $ \cP_\theta$ (see Sj{\"o}strand \cite{Sjostrand1997}) satisfies the following estimate from  \cite{BonyHafner2008}, Lemma 2.2:
  
{\em  There exists $\epsilon >0$ such that for   all $E'\in [a,b] +i[-ch,ch]$ and $\theta=Nh$ with $N>0$ large enough, 
\[\lb{2.27}\|(\cP_\theta-E)^{-1}\|_{\cL\left(\cH^0,\cH^{0} \right)}={\mathcal O}\left(h^{-M}\right)  \prod_{\small\begin{array}{l}z_j\in {\rm Res}\,({\cP_h})\\ | E-z_j^2|<{\epsilon\theta}\end{array}} \frac{h}{| E-z_j^2|},\qq |E-E'| <\epsilon\theta/2,\] where $M>$ is some number depending on $N.$ 

}

Now, from (\ref{2.25}), (\ref{2.26}), $\chi(P_h-E)^{-1}\chi= \chi (P_\theta-E)^{-1}\chi$ and from
the estimate 
\[\lb{lambdah}\langle\l\rangle \lesssim h^{-1}=l+1/2\lesssim\langle\l\rangle,\] we get (\ref{2.4}).

By Lemma \ref{l-interpol}, estimate (\ref{interpol}), we get for $j=1,2$
\[\lb{H2}\|\chi (\cP_h - E)^{-1}\chi f\|_{H^j(\R)}\leq C_jh^{-M} \langle E\rangle^{j/2} \prod_{\small\begin{array}{l}z_j\in {\rm Res}\,({\cP_h})\\ | E-z_j^2|<{\epsilon\theta}\end{array}} \frac{h}{| E-z_j^2|} \| f\|_{H^0(\R)}.\]

Now, applying (\ref{3.1}) to $\cD_h=-\s_3hD_x-\alpha\s_1$ and using (\ref{H2}) with $j=1$ and $z^2=E$ we get 
\begin{align} \|\chi({\cD_h}-z)^{-1}\chi\|_{\cL\left(\cH^0,\cH^{0} \right)}& \leq C\langle z\rangle\left(\|\chi_1(\cP_h^--E)^{-1}\chi_1\|_{\cL\left(H^0,H^{1} \right)}+\|\chi_2(\cP^+_h-E)^{-1}\chi_2\|_{\cL\left(H^0,H^{1} \right)}\right)\nonumber\\
&\leq Ch^{-M} \langle E\rangle \prod_{\small\begin{array}{l}z_j\in {\rm Res}\,({\cP_h})\\ | E-z_j^2|<{\epsilon\theta}\end{array}} \frac{h}{| E-z_j^2|}.\lb{prevD}
\end{align}

As $z=h\l$ and $\langle E\rangle $ is bounded (see (\ref{2.26})) we get $$ \|\chi({\cD_l}-\l)^{-1}\chi\|_{\cL\left(\cH^0,\cH^{0} \right)}= \| h\chi({\cD_h}-z)^{-1}\chi\|_{\cL\left(\cH^0,\cH^{0} \right)}\leq C \langle \l\rangle^C \prod_{\small\begin{array}{l}\l_j\in {\rm Res}\,({\cP_l})\\ |\l-\l_j|<1\end{array}} 
 \frac{1}{| \l-\l_j  |}$$
which is estimate  (\ref{2.4bis}).

Now, we discuss the location of the resonances.

From  \cite{Iantchenko2015} (see also Section \ref{Sect2}) we know that for any $N\in\N$  there exist     $r\in\N$ and functions $f_j=f_j(2k+1)={\mathcal O}\left( (2k+1)^j\right),$ $k\geq 0,$ $j=1,\ldots,r,$ polynomial in $2k+1$ of order $\leq j,$ such that 
there exists an injective map $b_N$ from the set of pseudopoles
 $$\G^N(h)=\left\{V_0(x_0)-\frac{ih}{2}\omega(2k+1)+\sum_{j=2}^r h^jf_j((2k+1);\,\,k=0,1,2,\ldots\right\},\qq \omega=\left(\frac12|V_0''(x_0)|\right)^\frac12,$$  
into the set of resonances ${\rm Res}\,(\cP_h)\setminus \{0\}= {\rm Res}\,(\cD_h)\cup{\rm Res}^{\rm S}\,(\cD_h),$ $$\cP_h=-h^2\partial^2_x + \alpha^2(x)\pm h\alpha'(x),\qq \cD_h=-\s_3hD_x-\alpha\s_1,$$ such that all the resonances in $$\Omega_\delta(h)=[a/2,2b]+i[-ch^\delta,ch^\delta]$$ are in the image of $b_N$ and for $b_N(\l)\in\Omega_\delta(h),$
$b_N(\l)-\l={\mathcal O}\left(|\l |^{-N}\right). $ Here $\delta >0$ is any constant.
In particular, the number of resonances of  $\cP_h$ and $\cD_h$ is bounded 
in $\Omega_\delta(h).$ Furthermore, the operators $\cP_h$ and $\cD_h$ have no resonances in
\[\lb{De}\Delta_\epsilon(h)=[a/2,2b]+i[-\epsilon h,ch]\] for some $\epsilon >0.$ 
Moreover, in \cite{BonyHafner2008} it was proven
that  $$ \|\chi (\cP_h - E)^{-1}\chi f\|_{\cL\left(H^0,H^0 \right)}\leq  C\frac{|\ln h|}{h}e^{C|\Im E | |\ln h|/h},\qq \mbox{for}\,\,E\in [a/2,2b]+i[-\epsilon h,ch],$$ which using (\ref{2.25}) and (\ref{lambdah})  implies (\ref{2.5}),
$$\|\chi(\cP_l-\l^2)^{-1}\chi\|_{\cL\left(H^0,H^0 \right)}\leq C\frac{\ln{\langle \l\rangle}}{\langle \l\rangle}e^{C|\Im\l |\ln\langle \l\rangle}. 
$$

Now, proceeding as before  in (\ref{H2}), (\ref{prevD}) we apply Lemma \ref{p-3.1} to    $\cD_h=-\s_3hD_x-\alpha\s_1$ with $z^2=E$ and use Lemma \ref{l-interpol}, (\ref{interpol}),  
 and get
\begin{align*} \|\chi({\cD_h}-z)^{-1}\chi\|_{\cL\left(\cH^0,\cH^{0} \right)}& <C\langle z\rangle\left(\|\chi_1(\cP_h^--E)^{-1}\chi_1\|_{\cL\left(H^0,H^{1} \right)}+\|\chi_2(\cP^+_h-E)^{-1}\chi_2\|_{\cL\left(H^0,H^{1} \right)}\right)\nonumber\\
&\leq C \langle E\rangle\frac{|\ln h|}{h}e^{C|\Im E | |\ln h|/h}.
\end{align*}
Then for $E\in\Delta_\epsilon(h)$ (see (\ref{De})) and using (\ref{lambdah}), $z=h\l,$ we get
(\ref{2.5bis}):
$$\|\chi(\cD_l-\l)^{-1}\chi\|_{\cL\left(\cH^0,\cH^0 \right)}= \| h\chi({\cD_h}-z)^{-1}\chi\|_{\cL\left(\cH^0,\cH^{0} \right)}\leq C\ln{\langle \l\rangle}e^{C|\Im\l |\ln\langle \l\rangle}. 
$$

(IV) Estimate (\ref{2.6}), 
$$ \|\chi(\cP_l -\l^2)^{-1}\chi\|_{\cL\left(H^0,H^0 \right)}\leq \frac{C}{\langle\l\rangle}e^{C|\Im\,\l|} ,$$
follows as in  \cite{BonyHafner2008}  for a similar operator.  As in that proof, for
$$\l\in [N,2N]+i[-C\ln N,C_0],$$ for some $C>0$ fixed and $N>>l,$ we define the new semi-classical parameter $h=N^{-1}$ and $$E=h^2\l^2\in h^2[N^2/2,4N^2]+ih^2[-4CN\ln N, 4C_0 N^{-1}]\,\,\subset\,\,[a,b]+i[-ch |\ln h|,ch],$$ for some $0<a<b$ and $0<c.$ Then, $\cP_l$ can be written as
$\cP_l-\l^2=h^{-2}(\tilde{\cP}_h-E),$ where
$$\tilde{\cP}_h=-h^2\partial^2_x +\mu \alpha^2(x)\pm\nu\alpha'(x),\qq\mu=h^2\left(l+\frac12\right)^2,\,\,\nu=h^2\left(l+\frac12\right).$$ For $N>>l,$ the coefficients $\mu,$ $\nu$ are small, and the operator $ \tilde{\cP}_h$ is uniformly non-trapping for $E\in [a,b].$ Then Lemma 2.4 from \cite{BonyHafner2008}  states that

{\em For all $\chi\in C_0^\infty(\R),$ there exist $\mu_0,$ $\nu_0,$ $h_0,$ $C$ $>0$ such that, for all $\mu <\mu_0,$ $\nu<\nu_0$ and $h<h_0,$  $\tilde{\cP}_h$ has no resonances in 
$[a,b]+i[-ch |\ln h|, ch].$ Moreover,
\[\lb{2.34} 
 \|\chi(\tilde{\cP}_h -E)^{-1}\chi\|_{\cL\left(H^0,H^0 \right)}\leq \frac{C}{h}e^{C|\Im\,E|/h}, 
\] for all $E\in [a,b]+i[-ch |\ln h|, ch].$

}

By Lemma \ref{l-interpol}, (\ref{interpol}), we get  for $j=1,2$ \[\lb{j1}\|\chi (\tilde{\cP}_h - E)^{-1}\chi f\|_{H^j(\R)}\leq C \frac{\langle E\rangle^{j/2}}{h}e^{C|\Im\,E|/h}\|f\|_{H^2(\R)}.\] Now, applying (\ref{3.1}) to $\tilde{\cD}_h=-\s_3hD_x-\mu^\frac12\alpha\s_1$ and using (\ref{j1}) with $j=1$ and $z^2=E$ we get 
\begin{align*} \|\chi({\cD_h}-z)^{-1}\chi\|_{\cL\left(\cH^0,\cH^{0} \right)}& \leq C\langle z\rangle\left(\|\chi_1(\tilde{\cP}_h^--E)^{-1}\chi_1\|_{\cL\left(H^0,H^{1} \right)}+\|\chi_2(\tilde{\cP}^+_h-E)^{-1}\chi_2\|_{\cL\left(H^0,H^{1} \right)}\right)\\
&\leq C \frac{\langle E\rangle}{h}e^{C|\Im\,E|/h}.
\end{align*}
As $z=h\l$ we get $$ \|\chi({\cD_l}-\l)^{-1}\chi\|_{\cL\left(\cH^0,\cH^{0} \right)}= \| h\chi({\cD_h}-z)^{-1}\chi\|_{\cL\left(\cH^0,\cH^{0} \right)}\leq C \langle E\rangle e^{C|\Im\,E|/h}\leq C'  e^{4C|\Im\,\l |}$$
which is estimate (\ref{2.6bis}).\hfill\BBox

\section{Resonance expansion.} \lb{s-Resexp}

Here we prove Theorem \ref{th-1.3}.

Recall that the
 energy is conserved along the evolution (\ref{consen}): $ 
\| e^{-it\cD}\psi_0\|_\cH= \|\psi_0\|_\cH.$
Then it follows that for $\Im\l >\epsilon >0$ the resolvent
$({\cD} -\l)^{-m}$ is bounded in  $\cH.$
Let $\mH^{-m}=({\cD} -\l)^m \cH$ and $$\|u\|_{\mH^{-m}}=\|({\cD} -\l)^{-m} u\|_{\cH}\leq C \|u\|_{\cH}.$$  

By orthogonal decomposition (\ref{dec1}) and (\ref{dec2}) we get also the spaces
$\mH_l^{-m}=({\cD_l} -\l)^m \cH^0,\,\,\cH^0=L^2(\R;\C^2);$  for $\Im\l >\epsilon >0$
$({\cD_l} -\l)^{-m}$ is bounded in $ \cH^0$ and \[\lb{newspace}\|u\|_{\mH_l^{-m}}=\|({\cD_l} -\l)^{-m} u\|_{\cH^0}\leq C \|u\|_{\cH^0}\] uniformly in $l.$

\no (i) Let ${\displaystyle \cD_l=-i\s_3\partial_x+q_l\s_1,\qq q_l=-\left(l+\frac12\right)\alpha(x).}$
Let $u\in L^2(\R,\C^2),$ then $({\cD_l}-\l)^{-1}u$ is   analytic function of $\l\in \C_+$ with values in $L^2(\R,\C^2),$ bounded for $\l\in \R+i\epsilon,$ $\epsilon >0.$ Therefore, $v(t)=v(x,t),$  solution of $$\left\{\begin{array}{rl} i\partial_tv(x,t)&={\cD_l} v(x,t)\\
v(x,0)&=u(x),\end{array}\right.$$ admits the spectral Fourier-Laplace representation
$$\theta(t)v(t)=\frac{1}{2\pi i}\int_\R e^{-i (\omega+i\nu) t}\left({\cD_l}-(\omega +i\nu)\right)^{-1}ud\omega,\qq \nu >0,$$ where $\theta(t)$ is the Heaviside function and the integral converges  in  the sense of distributions of $t\in\R$ with the values in $L^2.$

Equivalently, we write for all $t\geq 0$
\[\lb{3.8}e^{-it{\cD_l}}u=\frac{1}{2\pi i}\int_{-\infty +i\nu}^{\infty+i\nu}  e^{-i k t}\left({\cD_l}-k\right)^{-1}udk,\qq u\in L^2(\R,\C^2).\]
Henceforth, we denote by ${\cR}^l_\chi(k)$ the meromorphic extension of $\chi \left({\cD_l}-k\right)^{-1}\chi.$


 The following lemma is proved similar to Lemma 3.2 in \cite{BonyHafner2008}.
\begin{lemma}\lb{l-3.2} Choose $\chi\in C_0^\infty(\R)$ and $N\geq 0.$ There exist bounded operators $B_j\in\cL\left(\mH_l^{-m},\mH_l^{-m-1} \right),$ $j=0,\ldots,N_0,$ $m\in \N,$ and $B\in 
\cL\left(\mH_l^{-m},\mH_l^{-m-N-1} \right)$ such that
$${\cR}^l_\chi(k)=\sum_{j=0}^N\frac{1}{(k-i(\nu+1))^{j+1}}B_j +\frac{1}{(k-i(\nu+1))^{N+1}}B{\cR}^l_{\tilde{\chi}}(k)\chi,$$ for some $\tilde{\chi}\in C_0^\infty(\R)$ satisfying $\chi\tilde{\chi}=\chi.$
\end{lemma}

Following \cite{BonyHafner2008} we define
$$\tilde{\cR}_\chi^l(k)={\cR}^l_\chi(k)-\sum_{j=0}^1\frac{1}{(k-i(\nu+1))^{j+1}}B_j.$$ Then Lemma \ref{l-3.2} implies for $\Im\,k\leq\nu$
\[\lb{3.13} \|\tilde{\cR}_\chi^l(k)\|_{\cL\left(\mH_l^0,\mH_l^{-2} \right)}\lesssim\frac{1}{\langle k\rangle^2}
 \|{\cR}^l_\chi(k)\|_{\cL\left(\mH_l^0,\mH_l^{0} \right)}.\]
Using that \[\int_{-\infty +i\nu}^{\infty +i\nu}\frac{B_j}{(k-i(\nu+1))^{j+1}}e^{-ikt}d k=0\] in 
(\ref{3.8}) we get
$$\chi e^{-it{\cD_l}}\chi u=\frac{1}{2\pi i}\int_{-\infty +i\nu}^{\infty+i\nu}  e^{-i k t}\tilde{\cR}_\chi^l(k)udk,$$ where the previous integral is absolutely convergent in $\cL\left(\mH_l^0,\mH_l^{-2} \right).$

\begin{figure}[htbp]
\includegraphics[width =15 cm]{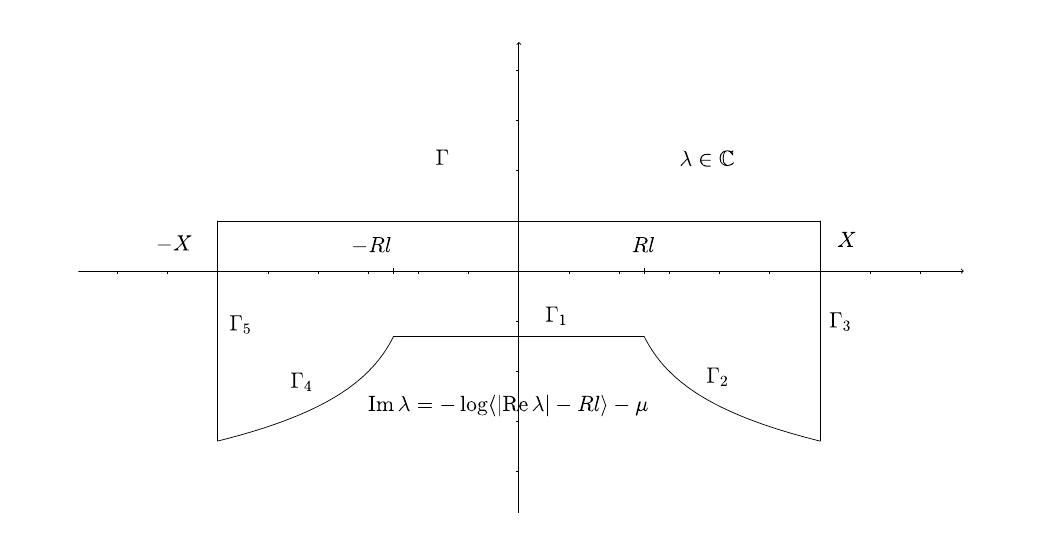}
 \caption{Integration paths from \cite{BonyHafner2008}.\label{Fig2} }
\end{figure}

Now, integrating along the path indicated in Figure \ref{Fig2} (reproducing Fig. 5 from \cite{BonyHafner2008})  we obtain by the Cauchy theorem
\[\lb{3.15} \frac{1}{2\pi i}\int_{-X +i\nu}^{X+i\nu}\!\!  e^{-i k t}\tilde{\cR}_\chi^l(k)udk=\!\!\!\!\sum_{\small\begin{array}{l}\l_j\in {\rm Res}\,({\cD_l})\\ \Im\l_j >-\mu\end{array}}\!\!\sum_{k=0}^{m(\l_j)-1}e^{-i\l_j t}t^k\pi_{j,k}^\chi u+\sum_{j=1}^5 \frac{1}{2\pi i}\int_{\Gamma_j}\!\! e^{-i \l t}\tilde{\cR}_\chi^l(\l)ud\l.
\]
Let $I_j=\int_{\Gamma_j}\!\! e^{-i \l t}\tilde{\cR}_\chi^l(\l)ud\l.$ 
Using (\ref{3.13}) and (\ref{2.6bis}): $  \|\chi({\cD_l}-\l)^{-1}\chi\|_{\cL\left(\cH^0,\cH^{0} \right)}\lesssim  e^{C|\Im\,\l|},$ we get
\begin{align} \|I_3\|_{\mH_l^{-2} }&\lesssim \int_{X-i\log\langle X\rangle}^{X+i\nu}\| e^{-i s t}\tilde{\cR}_\chi^l(s)u\|_{\mH_l^{-2} }ds\lesssim \int_{X-i\log\langle X\rangle}^{X+i\nu} e^{ t\Im s }\frac{1}{\langle s\rangle^2}\|{\cR}^l_\chi(s)u\|_{\cH^0 }ds\nonumber\\
&\lesssim\int_{X-i\log\langle X\rangle}^{X+i\nu} e^{ t\Im s+C|\Im s| }\frac{1}{\langle s\rangle^2}ds\|u\|_{\cH^0 }\lesssim\frac{1}{X^2}  \int_{-\log\langle X\rangle}^{\nu} e^{ t s +C|s|}ds\,\|u\|_{\cH^0}\nonumber\\& \lesssim\frac{1}{ X^2} \frac{e^{ t\nu}}{t}\,\|u\|_{\cH^0} \lb{3.16}
\end{align}
for $t>C.$ We now take the limit  in $X\rightarrow\infty$ in the $\cL\left(\mH_l^0,\mH_l^{-2}\right)$ sense in (\ref{3.15}). The integrals $I_3$ and $I_5$ go to $0$ thanks to  (\ref{3.16}).

We replace the paths $\Gamma_\bullet$ in the integrals  $I_2$ and $I_4$   by paths naturally extending  to $\infty.$ The extended paths are  denoted  again by $\Gamma_\bullet$. Note that 
\[\lb{3.17}\int_{\G_4\cup\G_1\cup\G_2}\frac{B_j}{(k-i(\nu +1))^{-j-1}}e^{-ikt}dk=0,\]
where the integral is absolutely convergent in $\cL\left(\mH_l^0,\mH_l^{-2}\right).$ On the other hand, using (\ref{2.4bis}) we have the following estimate, for $t$ large enough,
\begin{align}
\| I_1\|_{\cH^0}&\lesssim 
\int_{-Rl-i\mu}^{Rl-i\mu}\|e^{-i \l t} {\cR}^l_\chi(\l)u_0\|_{\cH^0}d\l\lesssim e^{-\mu  t}\int_{-R l}^{R l}\| {\cR}^l_\chi(s-i\mu)u_0\|_{\cH^0} ds\nonumber\\
&\lesssim e^{-\mu  t}\int_{-R l}^{R l} \langle s\rangle^Cds\,\, \| u_0\|_{\cH^0}\lesssim e^{-\mu  t}\, l^{C+1}\, \| u_0\|_{\cH^0}\lb{3.18}.\end{align}

 Now we use  (\ref{2.6bis}): 
$  \|\chi({\cD_l}-\l)^{-1}\chi\|_{\cL\left(\cH^0,\cH^{0} \right)}\lesssim e^{C|\Im\,\l|},$
for $\Re\l>Rl$ and $C_0\geq\Im\l\geq -C_0-C_1\log\langle\l\rangle.$ 
Therefore,
\begin{align}
\| I_2\|_{\cH^0}&\lesssim\int_0^\infty\|e^{-i(Rl+s-i(\mu+\log\langle s\rangle))t}{\cR}^l_\chi(Rl+s-i(\mu+\log\langle s\rangle))u\|_{\cH^0}d s\nonumber\\
&\lesssim\int_0^\infty e^{-\mu t} e^{-t\log{\langle s\rangle } } e^{C(\mu+\log\langle s)} ds\|u\|_{\cH^0}\lesssim e^{-\mu t}\int_0^\infty \langle s\rangle ^{C-t}ds\|u\|_{\cH^0}\lesssim e^{-\mu t}\|u\|_{\cH^0}\lb{3.19}
\end{align}
for $t$ large enough. Similar estimate holds for $I_4.$ Since all these estimates hold in $\cL(\cH^0,\cH^0),$ (\ref{3.18}) and (\ref{3.19}) give the estimate of the rest (\ref{1.12})   with $M=C+1.$


The estimate (\ref{1.13}) follows from (\ref{1.10})  and (\ref{2.4bis}).
\vspace{0.5cm}

\no (ii) We choose $0> -\mu >\sup\{ \Im\l;\,\,\l\in {\rm Res}(\cD_l)\}$ and the integration path as in part (i) of the proof. We first suppose $e^{\epsilon' t} >l+1/2$ for some $\epsilon' >0$ to be chosen later. Then estimate (\ref{3.18}) can be replaced by
$$\| I_1\|_{\cH^0}\lesssim  e^{\left( (C+1)\epsilon'-\mu  \right)t}\, \| u_0\|_{\cH^0}.$$ Now, we suppose $l+1/2\geq  e^{\epsilon' t}.$ On one hand we have the inequality 
$$\|\chi e^{-it\cD_l}\chi\|_{\cL(\cH^0,\cH^0)}\lesssim 1.$$
On the other hand by hypothesis on $g$ we have
$$1\leq \frac{g\left(e^{\epsilon' t}\right)}{g(l+1/2)},$$
as $g$ is strictly decreasing.  It follows that
$$\|\chi e^{-it\cD_l}\chi\|_{\cH^0}\lesssim \frac{g\left(e^{\epsilon' t}\right)}{g(l+1/2)}.$$ This concludes the proof of the theorem if we choose $\epsilon$ sufficiently small and put $\epsilon:=\min\{\epsilon',\mu-(C+1)\epsilon'\}.$



\end{document}